\documentclass[twocolumn]{IEEEtran}

\usepackage{epsfig}

\usepackage{}

\def\BibTeX{{\rm B\kern-.05em{\sc i\kern-.025em b}\kern-.08em
    T\kern-.1667em\lower.7ex\hbox{E}\kern-.125emX}}

\newcommand{\E}{\mbox{E}}

\begin{document}
\title{Uplink Throughput in a Single-Macrocell/Single-Microcell CDMA System, with Application to Data Access Points}
\author{Shalinee Kishore, Stuart C. Schwartz, Larry J. Greenstein, H. Vincent Poor\thanks{S. Kishore is with the Department of Electrical
and Computer Engineering, Lehigh University, Bethlehem, PA, USA.
H.V. Poor and S.C. Schwartz are with the Department of Electrical
Engineering, Princeton University, Princeton, NJ, USA.
L.J. Greenstein is with WINLAB, Rutgers University, Piscataway, NJ, USA.  This research was jointly supported
by the New Jersey Commission on Science and Technology, the
National Science Foundation under Grant CCR-00-86017, and the
AT\&T Labs Fellowship Program. Contact:
skishore@eecs.lehigh.edu.}}

\maketitle

\begin{abstract}
This paper studies a two-tier CDMA system in which the microcell base is converted into a {\em data access point} (DAP), i.e., a limited-range base station that provides high-speed access to
one user at a time. The microcell (or DAP) user operates on the same
frequency as the macrocell users and has the same chip rate.  However,
it adapts its spreading factor, and thus its data rate, in
accordance with interference conditions.  By contrast, the
macrocell serves multiple simultaneous data users, each with the same
fixed rate.  The achieveable throughput for individual microcell
users is examined and a simple, accurate approximation for its
probability distribution is presented.  Computations for average throughputs, both
per-user and total, are also presented.  The numerical results highlight the impact of a {\em desensitivity} 
parameter used in the base-selection process.

\end{abstract}

\section{Introduction}
\label{dap_sec} 
In \cite{skishore1}, we studied the uplink user capacity in a two-tier CDMA
system composed of a single macrocell within which a single microcell is embedded.  This system assumed that all users transmit over
the same set of frequencies, with handoff between tiers.
Analytical techniques were developed to both exactly compute and
accurately approximate the user capacity supported in such a
two-tier system. Here, we implement these techniques to study a
particular kind of single-macrocell/single-microcell system. Specifically, we
consider a wireless data system where the microcell is designed to attract only a
small number of users $(n)$, while the macrocell attracts a
larger number of users $(N_M)$. The users are uniformly
distributed over the entire system coverage area, all having the same chip rate, $1/W$, where $W$ is the system bandwidth.
The macrocell users have the same fixed data rate, $R_M$, and
the same output signal-to-interference-plus-noise ratio (SINR)
requirement, $\Gamma_M$. The $n$ microcell users, on the other
hand, can be high-speed and are given access to the
microcell one-at-a-time.  The
microcell resembles a {\em data access point} (DAP), i.e., a base
station with limited coverage area that provides high data rates
to a small number of users in sequence.  Some examples of DAP-like applications are in \cite{Kohno}-\cite{Goodman}.
The uplink data rate, $R_\mu$, of any given microcell user has a
maximum achieveable value, $R_\mu^*$, determined by existing
interference conditions. We will quantify $R_\mu^*$ and related
quantities for a given a microcell
output SINR requirement.  We focus here on the uplink, which we have
shown in \cite{sk_downlink} to be the limiting direction in this kind
of architecture.
\\
\\
In Section \ref{sysarch}, we elaborate on the system architecture,
the uplink SINR's, and the path gain model used.  In the process, we
identify a normalized desensitivity factor, $\zeta$, that is a key design
variable.  In Section \ref{datarates}, we derive the uplink transmission
rate for a given user and propose (and confirm via simulation) a simple approximation for its
probability distribution.  In Section \ref{throughputs}, we compute, as functions of
$\zeta$, the average per-user and per-DAP throughputs and show that the choice of $\zeta$ involves a tradeoff between user
high-speed capability and overall DAP utilization.  We also show the tradeoff between DAP performance and total user population.
\section{System and Channel Assumptions}
\label{sysarch}
\subsection{Architecture}
\begin{figure}[htp]
\begin{center}
\epsfig{figure=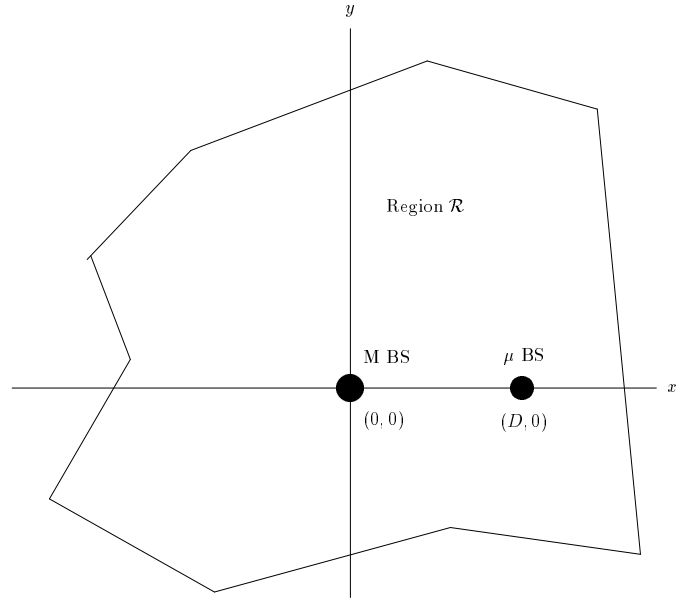,width=3.5in}
\caption{An example of a region $\mathcal R$, with macrocell and microcell
base station.  Here M BS and $\mu$ BS denote the macrocell and
microcell base stations, respectively.}\label{sysfig}
\end{center}
\end{figure}

We assume a system in which $N$ data users in some region $\mathcal R$
communicate with either a low-speed macrocell base or a microcell base
acting as a DAP, as shown in Fig. \ref{sysfig}.  In our computations, we will
further assume that the region $\mathcal R$ is a square of side $L$, with
the macrocell base at its center, the microcell base at some distance
$D$ from the center, and users located with uniform randomness over
the region.  
\\
\\
Each data user communicates with the macrocell at a fixed rate
$R_M=W/G$, where $G$ is the spreading factor of the CDMA code.
However, when a user's path gain to the microcell base (DAP) exceeds
some threshold, it becomes a candidate for communication, at higher
speed, with the DAP.  The base selection criterion we use is that the
macrocell base is selected whenever the user's path gain to it
exceeds that to the DAP by some fraction $\delta$, called the {\em
desensitivity factor}, or just {\em desensitivity}.  Clearly, the
smaller $\delta$ is, the smaller the number of users eligible for DAP
access. 
\\
\\
We denote by $n$ the number of eligible DAP users at any time, and
specify that they access the DAP sequentially.  Each gets a timeslot
whose duty cycle is $1/n$, and a data rate during the timeslot that
depends on factors discussed below.  The remaining $N_M=N-n$ users
simultaneously access the macrocell base at data rate $R_M$. These users require a
minimum SINR of $\Gamma_M$ while each of the $n$ DAP users requires a
minimum SINR of $\Gamma_\mu$.
\subsection{Uplink SINR's}
Each of the $n$ microcells users accesses the microcell one-at-a-time so that, at any one time, there are exactly $N_M+1$ system users. The base stations
employ matched filter (RAKE) receivers to detect these
active users.  Each base controls the transmit powers of its own
users so that each user meets the minimum output SINR requirement at
that base.  For such a system, the output SINR's at the macrocell and
microcell are
\begin{equation}
\mbox{SINR}_M = \frac{\frac{W}{R_M}S_M}{(N_M-1)S_M + S_\mu I_M +
\eta W} \label{sinr1}
\end{equation}
and
\begin{equation} \mbox{SINR}_\mu = \frac{\frac{W}{R_\mu}S_\mu}{S_M I_\mu +
\eta W},\label{sinr2}
\end{equation}
respectively, where $S_M$ and $S_\mu$ are the received power from each macrocell
user at the macrocell base and the received power from the
microcell user at the microcell base, respectively; $\eta W$ is
the received noise power, at each base, in bandwidth $W$; $R_M$
and $R_\mu$ are the bit rates of macrocell and microcell users,
respectively; and $I_M$ and $I_\mu$ are the normalized cross-tier
interferences.  Specifically,
$I_\mu$ is the normalized interference from the $N_M$ macrocell
users into the microcell base, and $I_M$ is the normalized
interference from the one active microcell user into the macrocell
base.  As shown in \cite{skishore1},
$I_\mu$ and $I_M$ depend solely on the set of path gains from the
active users to both bases, and on $\delta$.  They are
\begin{equation}
I_M=\frac{T_{M}}{T_{\mu}}
\label{im}
\end{equation}
and
\begin{equation}
I_\mu = \sum_{k \in M} \frac{T_{\mu k}}{T_{Mk}},
\label{imu}
\end{equation}
where $T_M$ ($T_{\mu}$) is the path gain from the microcell user to
the macrocell (microcell) base; $T_{\mu k}$ ($T_{Mk}$) is the path
gain from the $k$-th macrocell user to the microcell (macrocell) base;
and $M$ denotes the set of all macrocell users.  The denominators in
(\ref{im}) and in each term of (\ref{imu}) are a consequence of using
power control to each user from its serving base.  Adequate
performance requires that $\mbox{SINR}_M \geq \Gamma_M$ and
$\mbox{SINR}_\mu \geq \Gamma_\mu$.  
\subsection{Path Gain Model}
The path gain, $T$, between either base and a user at a distance $d$
is assumed to be
\begin{eqnarray}
T=\left\{ \begin{array}{cc} H\left( \frac{b}{d} \right)^2
10^{\chi/10}, & d \leq b \\ H \left( \frac{b}{d} \right)^4
10^{\chi/10}, & d>b \end{array} \right.,
\end{eqnarray}
where $b$ is the ``breakpoint distance'' (in the same units as $d$),
at which the slope of the dB path gain versus distance changes;
$\chi$ is a zero-mean Gaussian random variable for each user
position, with standard deviation $\sigma$; and $H$ is a proportionality
constant that depends on wavelength, antenna heights and antenna gains.
Note that $T$ is a local spatial average, so that multipath effects
are averaged out.  There can be different values of $b$ for the
microcell and macrocell, and similarly for $\sigma$ and $H$.  The
factor $10^{\chi/10}$ is often referred to as lognormal shadow fading,
which varies slowly over the terrain.  Both $\chi$ and $d$ are random
variables for a randomly selected user.   
\\
\\
Due to the greater height
and gain of the typical macrocell antenna, we can assume $H_M>H_\mu$.  As mentioned above, the
system we examine here assumes path-gain-based selection, i.e., a user
selects the macrocell base if the path gain to it exceeds the path
gain to the microcell base by some specified fraction $\delta$. In addition, the path gain of a user to the
macrocell exceeds its path gain to the microcell by a value
$h=H_M/H_\mu$ for the same distance and shadow fading. Therefore,
base selection
depends fundamentally on the ratio $\delta/h$, as opposed to $\delta$ alone.
(We chose not to normalize $\delta$ by $h$ in our study in \cite{skishore1}.) The factor $\delta/h$ is called the {\em normalized
desensitivity} of the microcell, and is hereafter denoted by
$\zeta$. Smaller values of $\zeta$ correspond to higher path gain
requirements to the microcell. This means that outlying users are
generally excluded from access to the microcell, and a smaller microcell
coverage area results.
\section{Data Rate for a DAP User}
\label{datarates}
\begin{figure}[htp]
\begin{center}
\epsfig{figure=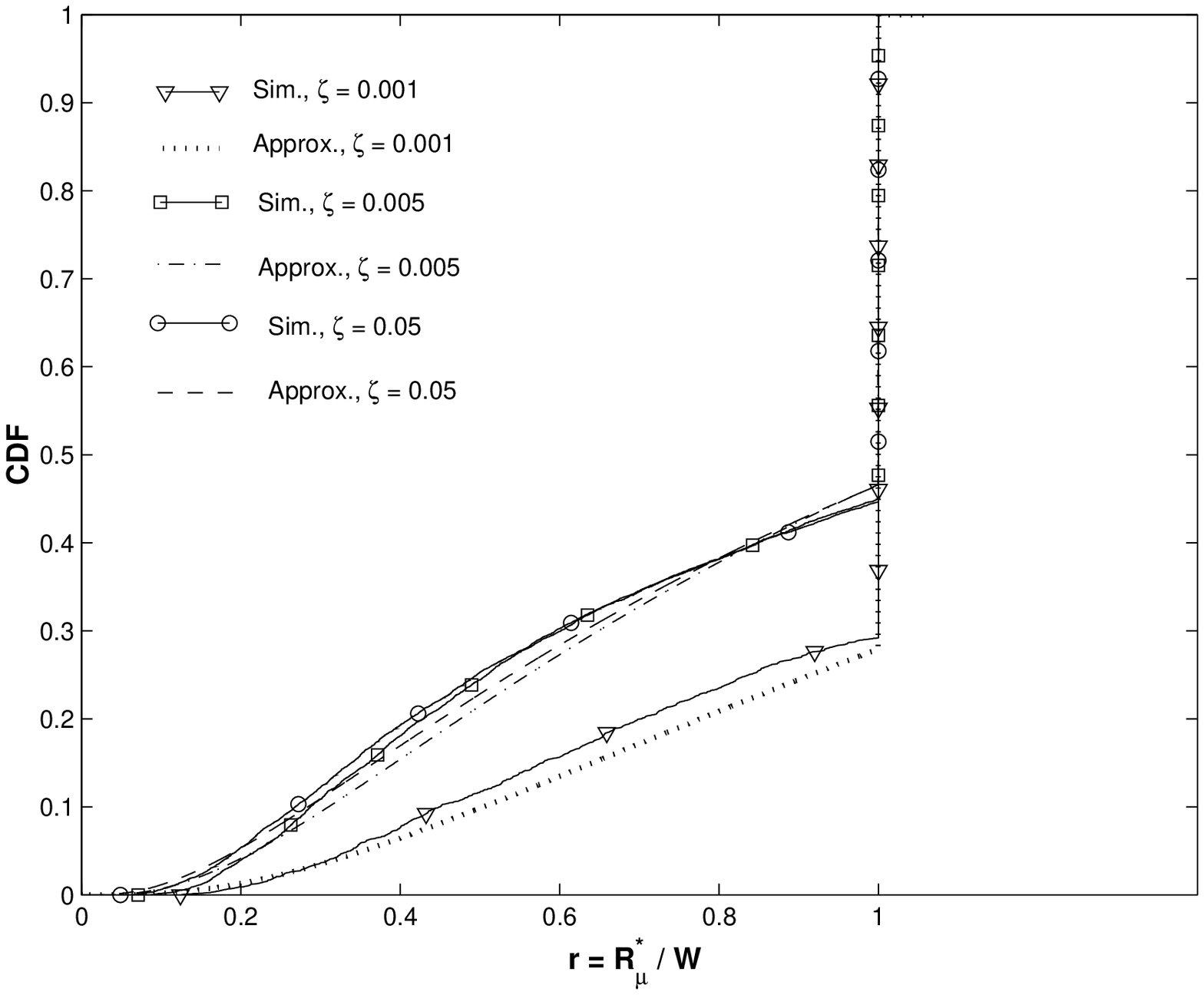,width=3.5in}
\caption{CDF's of $r$ using both simulation and the approximation
method, for $\zeta = 0.001$, $0.005$ and $0.05$. Results are for $N=26$.} \label{cdfs_singleuser}
\end{center}
\end{figure}

For a given number of DAP users $n$, we have
$N_M=N-n$ macrocell users, and there is a probability $p_n$ (derivable from the
results in \cite{skishore1}) that $n$ of the $N$ users will choose
the microcell and $N-n$ the macrocell.  We can show from (1) and (2) that the requirements $\mbox{SINR}_M \geq \Gamma_M$ and $\mbox{SINR}_\mu \geq \Gamma_\mu$ yield
positive solutions for $S_\mu$ and $S_M$ if and only if
\begin{equation}
\frac{W}{R_\mu}(K-N_M) - \Gamma_\mu I_M I_\mu \geq 0, \label{feas1}
\end{equation}
where $K=W/(R_M \Gamma_M)+1$ is the {\em single-cell pole capacity} of the
macrocell \cite{gilhousen}.  Thus, we can find $R_\mu^*(n)$, the maximum achieveable
$R_\mu$ when $n$ users choose the microcell:  Using (\ref{feas1})
with $N_M = N-n$, and noting that the spreading factor
$W/R_\mu$ cannot be less than 1, we have
\begin{equation}
R_\mu^*(n)=\min \left( \frac{W(K-N+n)}{\Gamma_\mu I_M I_\mu},W
\right) , \mbox{~~~}1 \leq n \leq N.\label{rmueq}\footnote{We consider all integer
spreading factors from 1 to $W/R$.  For analytical
ease, we further assume a {\em continuum} of values over that range, so
that $R^*_\mu$ is the value of $R_\mu$ that meets (6) with equality,
up to a maximum of $W$.}
\end{equation}
We define $r=R_\mu^*/W$ and, henceforth, we examine this normalized data rate. Note that $r$ will be different as each of the $n$ users
takes its turn, because $I_M I_\mu$ depends on the set of all (active)
user path gains.\footnote{Since this set depends on $\zeta$, the
product $I_MI_\mu$ in (7) does, as well.  The denominator $T_\mu$ in
(3), which decreases with increasing microcell coverage area (i.e.,
with increasing $\zeta$), has the dominant impact on $I_MI_\mu$ and,
thus, $R^*_\mu$.}  Also, for any given user, we see that $I_M$ and
$I_\mu$ are random variables because of the random locations and
shadow fadings of all users.  Thus, $r$ is itself a random
variable. To facilitate analysis, we assume (and will show) that
$I_M$ and $I_\mu$ can be treated as lognormal variates, whose first
and second moments are
obtainable using the results in 
\cite{skishore2}.  Since $1/(I_MI_\mu)$ is lognormal under this
assumption, we conclude that $r$, as given above,
is a truncated lognormal random variable.  The Appendix shows how the
lognormal parameters are obtained from the moments
of $I_M$ and $I_\mu$.
\\
\\
Let $F(r|n)$ denote the cumulative distribution function
(CDF) of $r$ for a given $n$.  The CDF of $r$ with the condition on
$n$ removed, but subject to $n$ exceeding 0, is\footnote{If $n=0$
(which happens with probability $p_0$), we have no interest in the
data rate of a non-existent user!  Thus, we seek the CDF of $r$ when
there is at least one DAP user.}
\begin{equation}
F(r)=\frac{\sum_{n=1}^N p_n F(r|n) \label{cdfsum}}{1-p_0}.
\label{cdfRmu}
\end{equation}
Assuming that $F(r|n)$ can be approximated as the CDF of a truncated lognormal random variable, $F(r)$ can approximated by a weighted
sum of CDF's of truncated lognormals.
\\
\\
To test the reliability of the lognormal approximation, we
performed a series of simulation trials.  We assumed a square
region $\mathcal R$ with sides of length $L$ over which users are
uniformly distributed, with the two bases separated by $D$. For the system and propagation
parameters listed in Table 1, we determined the achieveable
data rates of microcell users over 10,000 trials, for
various values of $\zeta$ and with $N=26$.\footnote{The case $N=26$
corresponds, for the parameters of Table 1, to a highly stressed
macrocell, i.e., $N \approx K$, the macrocell pole capacity.  Later we show the general tradeoff between $N$ and DAP
throughput.}  The resulting CDF of
$r$ is plotted in Fig. \ref{cdfs_singleuser}.
Along with simulation results, this figure contains CDFs obtained from
analysis, (\ref{cdfRmu}), assuming lognormal
cross-tier interferences.  These results show that the
lognormal approximation is reliable over a wide range of $\zeta$.
As $\zeta$ gets smaller, the data rates of the microcell users
increase, suggesting that smaller $\zeta$ is desirable.  The problem with {\em extremely}
small $\zeta$, however, is that it shrinks the population
of possible microcell users.  The result is under-utilization of the
DAP, as discussed below.
\section{Results}
\label{throughputs}
\subsection{Per-User and Total DAP Throughputs}
\begin{figure}[htp]
\begin{center}
\epsfig{figure=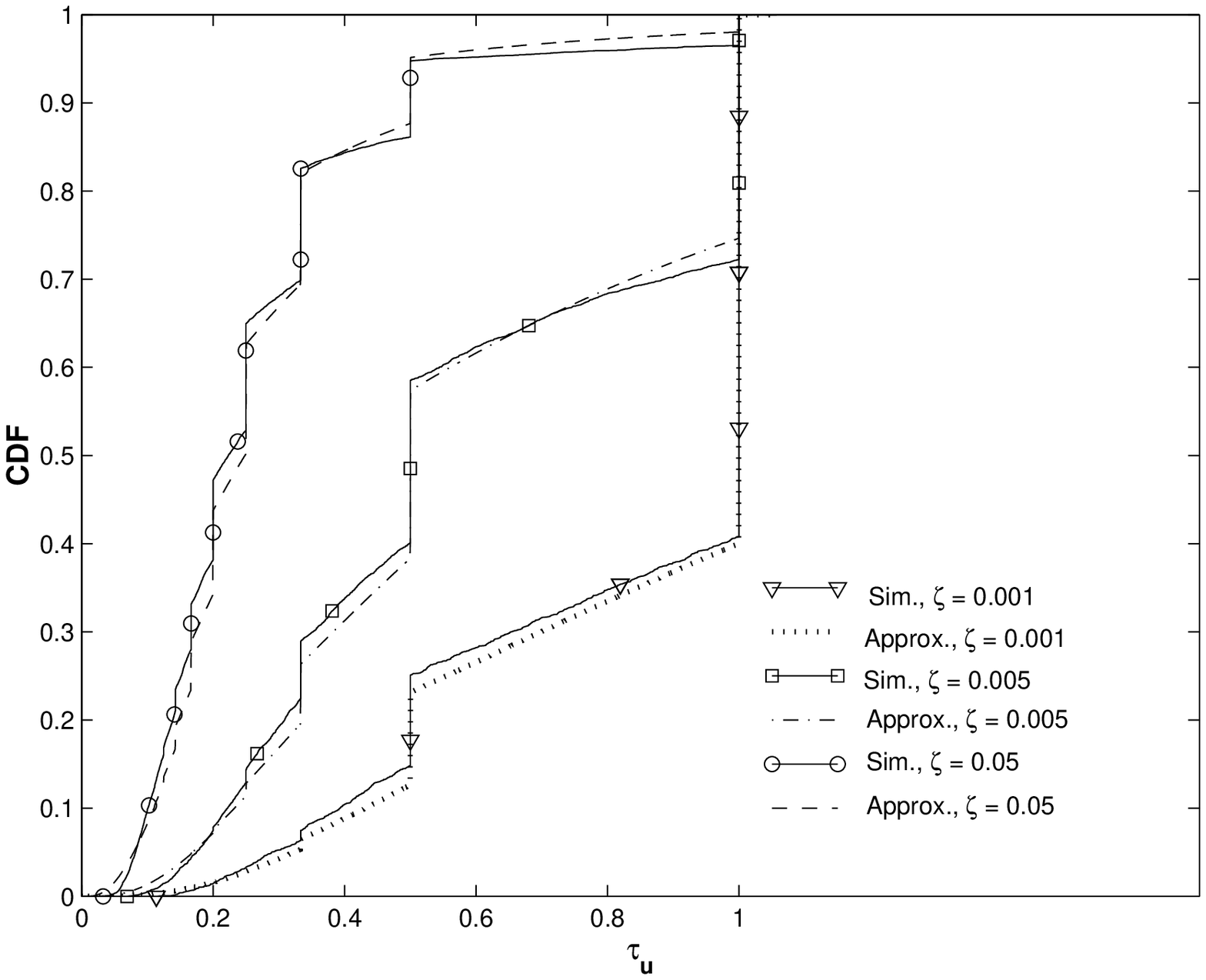,width=3.5in}
\caption{CDF of $\tau_u$ using both simulation and the approximation
method, for $\zeta = 0.001$, $0.005$ and $0.05$.  Results are for $N=26$.} \label{peruser_cdfs}
\end{center}
\end{figure}

\begin{figure}[htp]
\begin{center}
\epsfig{figure=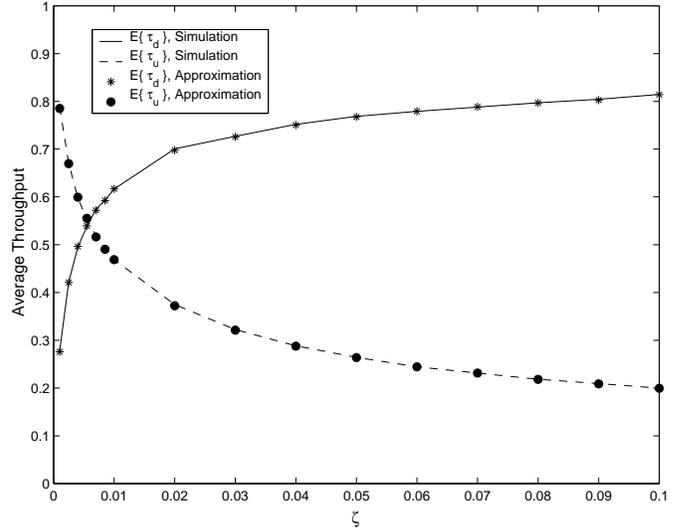,width=3.5in}
\caption{$\mbox{E}\{ \tau_u \}$ and $\mbox{E}\{ \tau_d \}$ as functions of
$\zeta$ using both simulation and the approximation method.  Results
are for $N=26$.}
\label{avg_thruputs}
\end{center}
\end{figure}

We define our throughput variable as the time-averaged data rate normalized by $W$.
The per-user throughput, $\tau_u$, takes into account the time-limited
access that data users must accept if there is more than one microcell
user in the system.  This division of time effectively reduces data
rates $(r)$ in a system with $n$ microcell users by a factor of $n$.
The CDF of $\tau_u$ is thus 
\begin{equation}
F_{\tau_u}(\tau_u)=\frac{\sum_{n=1}^N p_n F(n \tau_u|n)}{1-p_0}, \label{cdfsum2}
\end{equation}
where $F(\cdot|n)$ is as defined earlier. The CDF of $\tau_u$ can be approximated using the truncated-lognormal
assumption for $R_\mu^*$, (\ref{rmueq}).  Using both simulation
and this approximation, we again considered the
single-macrocell/single-microcell system described above, with the parameters in
Table 1.  The results are plotted in Fig.
\ref{peruser_cdfs} for three different values
of $\zeta$.  We see that the approximate distribution follows the
simulation distribution very closely for all three values.  The jumps
in Fig. 3 are related to the fact that per-user data rates are
confined to the discrete values $W/n$, $n=1,2,\ldots$ for the high
fraction of cases where $R^*_\mu=W$.
\\
\\
The {\em total} throughput for the DAP, for given a $n \geq 0$, is the sum of the
throughputs for the $n$ users.  We denote the total throughput (i.e., DAP utilization) by
$\tau_d$.  Thus, $\tau_d$ represents performance from the network operator's point-of-view, while $\tau_u$ represents performance from the user's point-of-view.  We quantify the average per-user and per-DAP throughputs, as
functions of $\zeta$, where the averaging is over random locations
and shadow fadings of the users.  $\E \{ \tau_u \}$ is obtained
via the formula for its CDF, (\ref{cdfsum2}).  $\E \{ \tau_d \} $
is obtained by noting that $\E \{ \tau_d | n \}$ is just $\E \{ r|n \}$.  Removing the condition on $n$, we get
\begin{equation}
\E \{ \tau_d \} = \sum_{n=0}^N p_n \E \{r|n \}.
\end{equation}
The results for $\E \{ \tau_u \}$ and $\E \{ \tau_d \}$ are shown in Fig.
\ref{avg_thruputs} for $N=26$.  They are given for both simulation and the
approximation method, and we see very strong agreement.  For this system, we see that, when
$\zeta \approx 0.007$, $\mbox{E}\{ \tau_u \} = \mbox{E} \{\tau_d \} \equiv \tau^*$.   This
value of $\zeta$ (which we call $\zeta^*$) is desirable,
since it balances the throughput for both individual users
and the overall  DAP.  When $\zeta > \zeta^*$, individual user throughputs are
compromised for the sake of higher DAP utilization; when $\zeta < \zeta^*$, higher per-user throughputs are
obtained, but for an under-utilized DAP.

\subsection{Effect of $N$ on $\tau^*$}
The results reported thus far assumed $N=26$, i.e., $N \approx K$.  It is instructive to repeat the above exercise for a range of $N$, to show the tradeoff between total user population and DAP throughput.  Thus, we have computed $\E \{ \tau_u \}$ and $\E \{ \tau_d \}$ as functions of $\zeta$ for various $N$, still assuming a uniform distribution of users.  For each value of $N$, $\E \{ \tau_u \}$ decreases and $\E \{ \tau_d \}$ increases with increasing $\zeta$, just as in Fig. 4; however, the values of $\tau^*$ and $\zeta^*$ change with $N$.  This is shown in Fig. 5.
\\
\\
For a given $\zeta$, decreasing $N$ implies a shrinking number of DAP users, leading to a less utilized DAP.  Thus, $\zeta$ must be increased to maintain high DAP utility, i.e., $\zeta^*$ must increase as $N$ decreases below $K$.  At the same time, the cross-tier interference at the DAP shrinks, leading to increases in $\tau^*$.  In the region $N \leq K$, the average $n$ at $\zeta=\zeta^*$ is close to 1, so that $N_M \approx N-1$.  In the region $N>K$, meeting the feasibility condition $N_M < K$, (6), requires that $n>1$.  The result is a sharp decline in $\tau^*$ as $N$ increases from $K$, as foretold by the steepness of the $\tau^*$-curve at $N=26$.

\subsection{Effect of Dense User Distribution Around the DAP}
Using the methods presented here, throughput results can be
obtained for a system with a denser user distribution around the
DAP. We found that, as the density around the
DAP increases, both $\zeta^*$ and $\tau^*$ decrease.  Smaller values of $\zeta^*$ are desirable because they help reduce $n$, the number of users that must share the DAP.  Despite this, as the user density around the DAP increases, the most probable value of $n$ increases, resulting in smaller $\tau_u$ and a net (modest) reduction in $\tau^*$.
\section{Conclusion}
We have demonstrated that by controlling the desensitivity factor, a
microcell can be converted to a data access point.  The data rate of
the single DAP user was analyzed and shown to be well-represented as a
truncated-lognormal variate.  A method was then developed to compute
the throughput statistics, and we found values of desensitivity, over the practical range of user population, for which both per-user and total DAP throughputs are reasonably high. 

\appendix
\section{Lognormal Representation of the User Data Rate}
The user data rate, $r=R_{\mu}^*/W$ is seen from (\ref{rmueq}) to be
the lesser of 1 and $Z$, where $Z=(K-N-n)/(\Gamma_\mu I_M I_\mu)$.  Defining $z = \ln Z$, we have $z = \ln ((K-N-n)/\Gamma_\mu) - \ln I_M - \ln I_\mu$.
Clearly, if $I_M$ and $I_\mu$ are lognormal, $z$ is a Gaussian random
variable whose mean and standard deviation are obtainable from the
means and standard deviations of $\ln I_M$ and $\ln I_\mu$.  
\\
\\
In \cite{skishore2}, the first and second moments of $I_M$ and $I_\mu$
are derived for the two-cell system.  The means and standard
deviations of $\ln I_M$ and $\ln I_\mu$ are then obtainable as
follows:  Let the mean and standard deviation of $\ln I_M$ be $m$ and
$\sigma$, respectively.  From lognormal statistics \cite{andrealarry},
\cite{schwartzye},
\begin{eqnarray*}
\E \{ I_M \} = e^{m}e^{\sigma^2/2} \mbox{~~~and~~~~}
\E\{ I_M^2 \} = e^{2m}e^{2 \sigma^2}.
\end{eqnarray*}
Solving for $m$ and $\sigma$,
\begin{eqnarray*}
m= \frac{1}{2} \ln \left(\frac{(\E \{ I_M \})^4}{\E \{ I_M^2 \} } \right)\mbox{~~;~~~}
\sigma = \sqrt{\ln \left( \frac{ \E \{ I_M^2 \} }{(\E \{ I_M \})^2} \right)}.
\end{eqnarray*}
Similar results apply to $\ln
I_\mu$.  Thus, $r$ can be fully described as a truncated lognormal
variate given the first two moments of $I_M$ and $I_\mu$, and assuming
both are lognormal.

\vspace{.2in}
\begin{table}[h]
\begin{center}
\begin{tabular}{|c|c||c|c|}
\hline $W/R_M$ & 128 & $L$ & 1 km
\\ $\Gamma_M$ & 7 dB & $\Gamma_\mu$ & 8.45 dB
\\ $b_M$ & 100 m & $b_\mu$ & 100 m
\\ $H_M$ & $10 H_\mu$ & $D$ & 300 m
\\ $\sigma_M$ & 8 dB & $\sigma_\mu$ & 4 dB
\\
\hline
\end{tabular}
\\
\caption{System Parameters Used}
\end{center}
\label{sys_param}
\end{table}
\begin{figure}[htp]
\begin{center}
\epsfig{figure=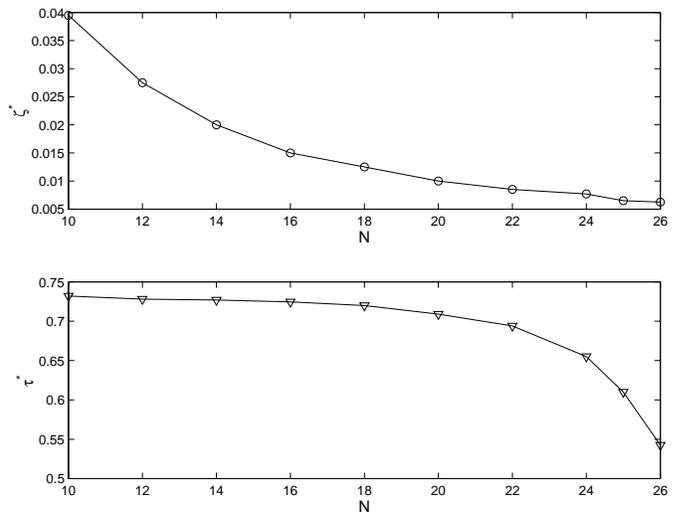,width=3.5in}
\caption{$\zeta^*$ and $\tau^*$ as functions of $N$, for $10 \leq N \leq K$.}\label{densepop}
\end{center}
\end{figure}
\end{document}